# Graphene-based Nanoscale version of da Vinci's Reciprocal Structures

*Alexandre F. Fonseca*[*] *and Douglas S. Galvão*

Applied Physics Department, Institute of Physics "Gleb Wataghin", State University of Campinas, Campinas, SP, 13083-970, Brazil.

ABSTRACT: A reciprocal structure (RS) is a mechanical resistant structure formed by a set of self-supporting elements satisfying certain conditions of structural reciprocity (SR). The first condition is that each element of the structure has to support and be supported by the others. The second condition is that these functions cannot occur in the same part of the element. These two properties make beams and two-dimensional materials very much appropriate to build RSs. Commonly seen in floors or roofs, SR is also present in art, religious symbols, and decorative objects. Da Vinci has drawn several examples of such RSs. Here, thermal stability and mechanical resistance against impacts of simple nano versions of da Vinci's RSs based on graphene nanoribbons, were investigated through fully atomistic molecular dynamics (MD) simulations. We considered structures with three and four joins with and without RS topologies. Our MD results showed that 3-fold RSs are not thermally stable and that the 4-fold RSs can become thermally stable as long as the graphene nanoribbons have their external extremities fixed and either are not lengthy or have a kind of notch at the nanoribbons junctions. For these thermally stable structures, our results show that those with RS topologies are more impact resistant than those without SR, despite the fact that the used graphene nanoribbons are highly pliable. We discuss these results in terms of the number of joins, energy absorption, and stress on the structures. We discuss possible applications in nanoengineering.

---

[*] Corresponding author. Tel: + 55 19 35215364. E- mail: afonseca@ifi.unicamp.br (Alexandre F. Fonseca)

**1. Introduction**

Structural reciprocity (SR) is a concept of self-supporting of load-bearing bars that together form larger mechanical resistant structures [1]. Dating back to the Neolithic, SR was found from native tepees and tents to old bridges like the one over the Rhine that was built in the Roman Empire by Julius Caesar. The drawings of Leonardo da Vinci [2] show structures satisfying SR. Commonly seen in floors or roofs, SR is also present in art, religious symbols, and decorative objects. Although SR involves a mutual exchange of action and reaction between parts of the whole structure, it is also known to rely on a perfect symmetric relationship between them [1]. From now on, any structure having SR will be called a "reciprocal structure" and will be referred to as "RS". The main characteristics of an RS are, first, the role of supporting and being supported should not occur in the same part of the structure, i. e., they must be separated, not overlapping like in truss bars. Second, each element of an RS must, at the same time, support the others and being supported by the others. These two properties make beams and two-dimensional materials very much appropriate to build RSs. Figure 1 shows two examples of da Vinci's RSs with three and four-folds built with knives and glasses on the floor. These structures are also called reciprocal frames [3] and it is a matter of architectural applications worldwide.

Here, we present a study of some simple *nano* versions of two da Vinci's RSs based on graphene nanoribbons (structures shown in Figure 2). Thermal stability and mechanical resistance against impacts were investigated through fully atomistic molecular dynamics (MD) simulations. We considered two different structures with three and four joins, with and without RS topologies, made of graphene nanoribbons of two different length sizes and three different width sizes, for comparison. These structures will be called 3-fold or 4-fold. We also tested, for comparison, the mechanical resistance of a similar pristine graphene structure. Our MD results showed that structures based on graphene nanoribbons with long lengths are not stable under thermal fluctuations, even if their external extremities are fixed (the extremities far away from the center of the structures). The graphene pliability leads the full structure to collapse even for those with a large width. However, RS structures made of small length graphene nanoribbons and external extremities fixed are thermally stable. The effect of the presence of a small notch in the region where the nanoribbons contact each other was also considered. The notch was created by hydrogenation of the graphene nanoribbon around the region of contact with other nanoribbons. Notched RSs of 4-fold were, then, shown to be thermally stable. As the structures with three-fold were shown to always deform and/or collapse, the mechanical resistance of only 4-fold structures were analyzed. Then, we show that structures with RS topologies of 4-fold are more impact resistant than that without structural reciprocity, despite the fact that the used graphene nanoribbons are highly pliable. We discuss these results in terms of the flexure of graphene and possible applications in building self-sustained and resistant nano-domes and nanocages, as well as possible applications in nanoengineering.

**2. Models and computational methods**

*2.1. RS and non-RS structures*

In order to address the issue of mechanical resistance, we generated not only the 3- and 4-fold graphene-based RSs (left side of Figure 2) but also 3- and 4-fold graphene-based non-RSs as shown in the right side of Figure 2. The non-RSs were built by placing one or more graphene nanoribbons on top or bottom of the other nanoribbons, so breaking the rule of RSs of having each element of the structure supporting and being supported by the others. All graphene nanoribbons were Hydrogen passivated to avoid the formation of chemical bonds. The nanoribbons interact only via van der Waals interactions.

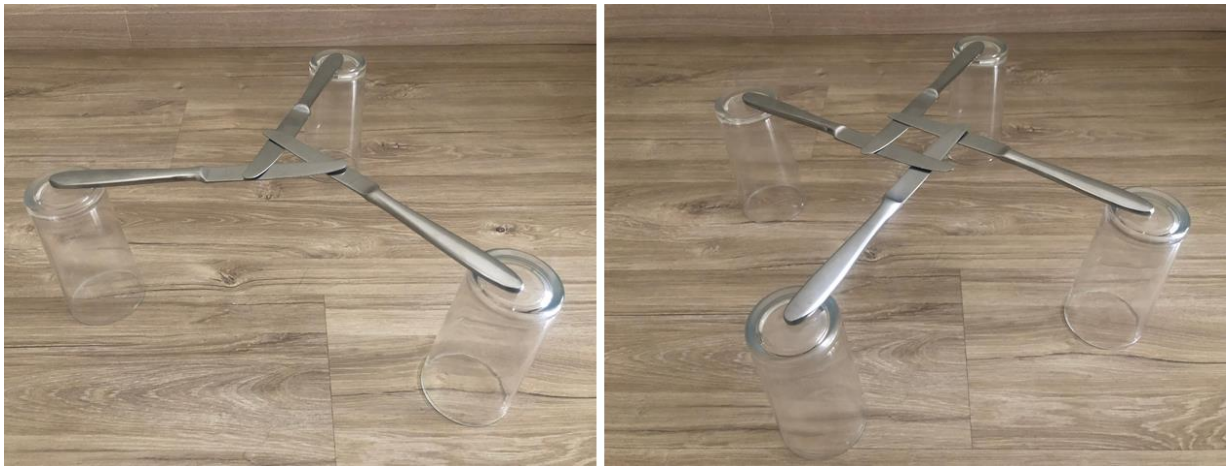

**Figure 1**: Simple realizations, using knives and glasses on the floor, of two examples of da Vinci's sketched RSs that will be studied here, with three (left) and four (right) folds.

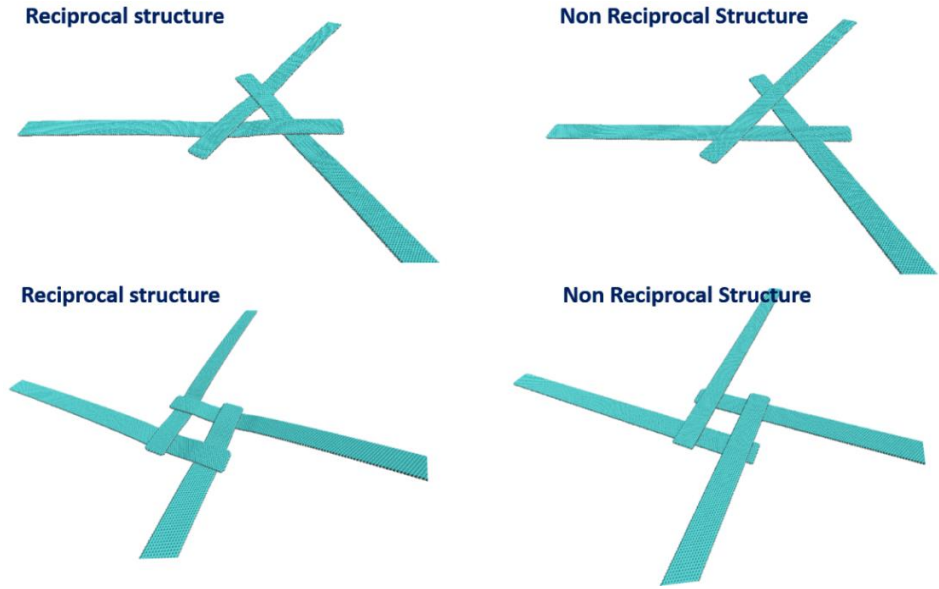

**Figure 2**: Graphene-based RSs (left column) and corresponding non-RSs (right column). In the non-RSs, one or more graphene nanoribbons are placed on top or bottom of the other nanoribbons.

The graphene nanoribbons used to build the RSs and non-RSs shown in Figure 2 were generated with two different values of lengths, ~ 165 and 400 Å, and three different values of widths, ~ 15.1, 25.2 and 50.4 Å. As some of these structures will be shown to be not thermally stable, we have generated a local "notch" by hydrogenating the regions around the contacts between different nanoribbons. Figure 3 shows one example for the 3- and 4-fold notched RSs.

*2.2. Computational methods*

The AIREBO [4] potential, as available in the LAMMPS [5] computational package, is used for the MD simulations. An energy minimization method is first applied to optimize the geometry of the structures (with force tolerance of $10^{-6}$ eV/Å), then MD simulations at 300 K are carried out for, at least, 2 nanoseconds using a Langevin thermostat. The time step of integration of the movement equations and thermostat damping factor were set in 0.5 fs and 1 ps, respectively. After these 2 ns of simulation at 300 K, the structures that hold the original configuration (RS or non-RS) will be considered as "thermally stable", and will be subjected to a mechanical test to investigate their rigidity and energy absorption against impact.

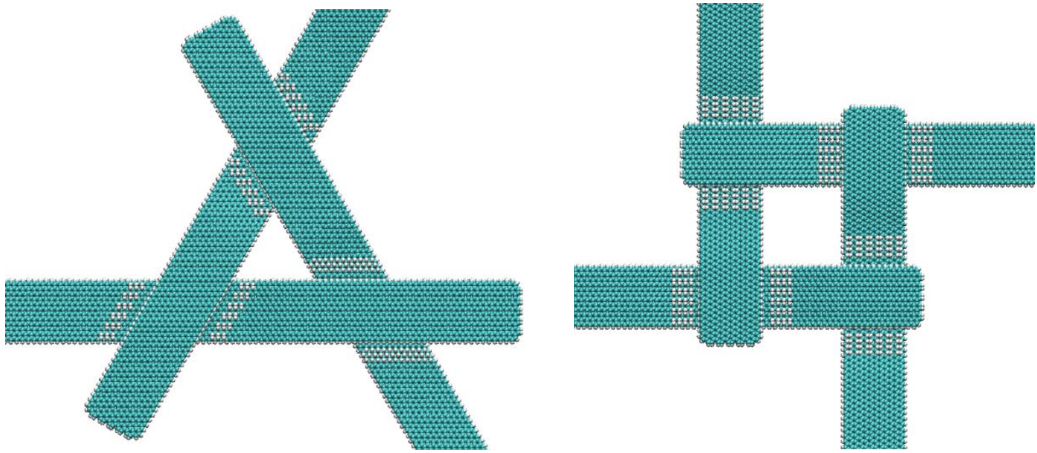

**Figure 3**: Top views of "notched" graphene-based RSs with 3-fold (left) and 4-fold (right). Carbon (Hydrogen) atoms are shown in cyan (white).

We have opted to use a quasi-static (instead of a dynamical one) approach for the impact tests. This method allows for a better estimation of the elastic energy of the system during the deformation it stands for the size and simulation time of the structures we are investigating here. It consists of placing a parallelepiped diamond projectile close to the plane of RSs or non-RSs initially at about 3.3 Å of distance, at their center. While diamond atoms are kept fixed, energy minimization is performed on the RS or non-RS structure. After that, the diamond projectile is displaced by 0.2 Å towards the perpendicular direction of the RS or non-RS structure and a new energy minimization is performed with the diamond atoms, again, held fixed. The extremities of the graphene nanoribbons are also kept fixed otherwise the whole structure will move along with the projectile. We repeat the above steps by a thousand times in order to simulate a quasi-statically collision between the projectile and the RSs or non-RSs. The total amount of displacement will be 200 Å and it is expected that the projectile breaks through the structure and, then, the total stress and energy of the RSs or non-RSs are obtained as functions of the $z$ coordinate ($z$ is the direction of displacements) of their center of mass. The parallelepiped diamond has planar lateral sizes of 100 Å and thickness of 30 Å (thickness is measured along RSs perpendicular direction). The extremities of all graphene nanoribbons were kept fixed during this impact mechanical test. AIREBO was considered to simulate the interaction between the diamond projectiles and the RSs or non-RSs.

The values of toughness and the differences of total energy before and after the mechanical test will be calculated for comparison between the RSs and non-RSs.

**3. Results**

*3.1. Thermally stable RSs*

The first results are concerned with the thermal stability and the equilibrium structures at room (300 K) temperature. The purpose here is to verify the effects of thermal fluctuations on the structure of the system. First, the thermal stability will be verified without the application of any external constraint. Then, structures with their external extremities fixed will be tested. By "external extremities fixed" we mean the extremities of the graphene nanoribbons that are far away from the RSs centers. As the graphene nanoribbons are very pliable, it was verified if any deformation caused by thermal fluctuations leads the RSs to change shape or configuration so becoming non-RS. The next four figures show these results for 3-fold and 4-fold RSs, and both the two different lengths and three different widths considered in this study. Tests were also performed on non-RSs but not shown because finding thermally stable RSs is one of the goals of this study. Figures 4 and 5 show snapshots of partially equilibrated 3-fold and 4-fold RSs, respectively, for the largest length of graphene nanoribbons (~ 400 Å) obtained after 100 ps and 2ns, also respectively, of MD simulations at 300 K.

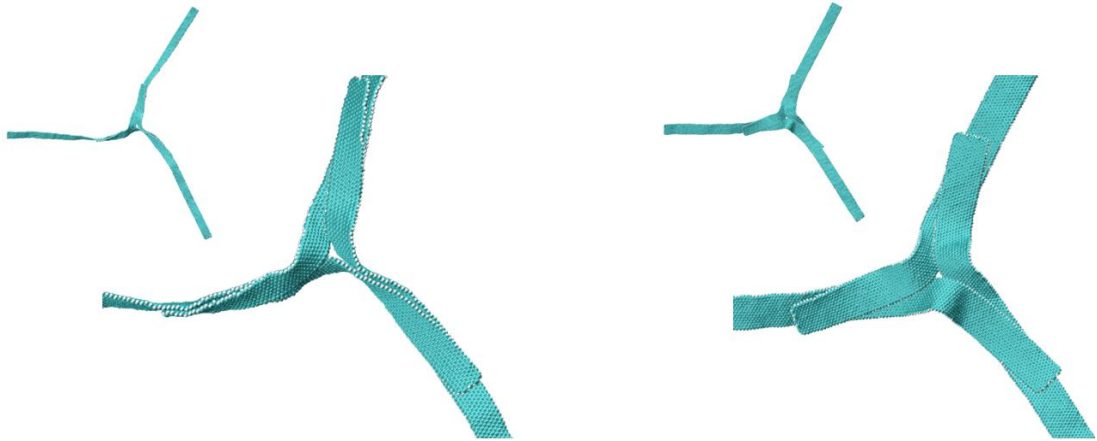

**Figure 4**: Partially equilibrated graphene-based 3-fold RSs made with graphene nanoribbons of 400 Å of length, and 15.1 (left) and 25.2 Å (right) of width, after 100 ps of MD simulations at 300 K. The four insets show the whole structures, and the larger figures show their central parts in order to provide an enlarged view. Cyan (white) colors represent carbon (hydrogen) atoms.

Figure 4 shows that the graphene nanoribbons at the center of the RSs, due to the maximization of the van der Waals interactions, are bent and twisted. For the 3-fold RS, it is not possible to say if the conditions for structural reciprocity continue to be satisfied. It is because the increased overlapping due to van der Waals attraction makes it not possible to say that each nanoribbon sustains and is sustained by the others at different locations. The snapshots shown in Figure 4 are not stable structures because the overlapping between the graphene nanoribbons surfaces keeps increasing along more time of MD simulations. Notice that it is enough to simulate the 3-fold RSs by about 100 ps to see that the SR might be compromised. For the 4-fold RSs, we had to simulate the structures for more than 10 times that of 3-fold ones.

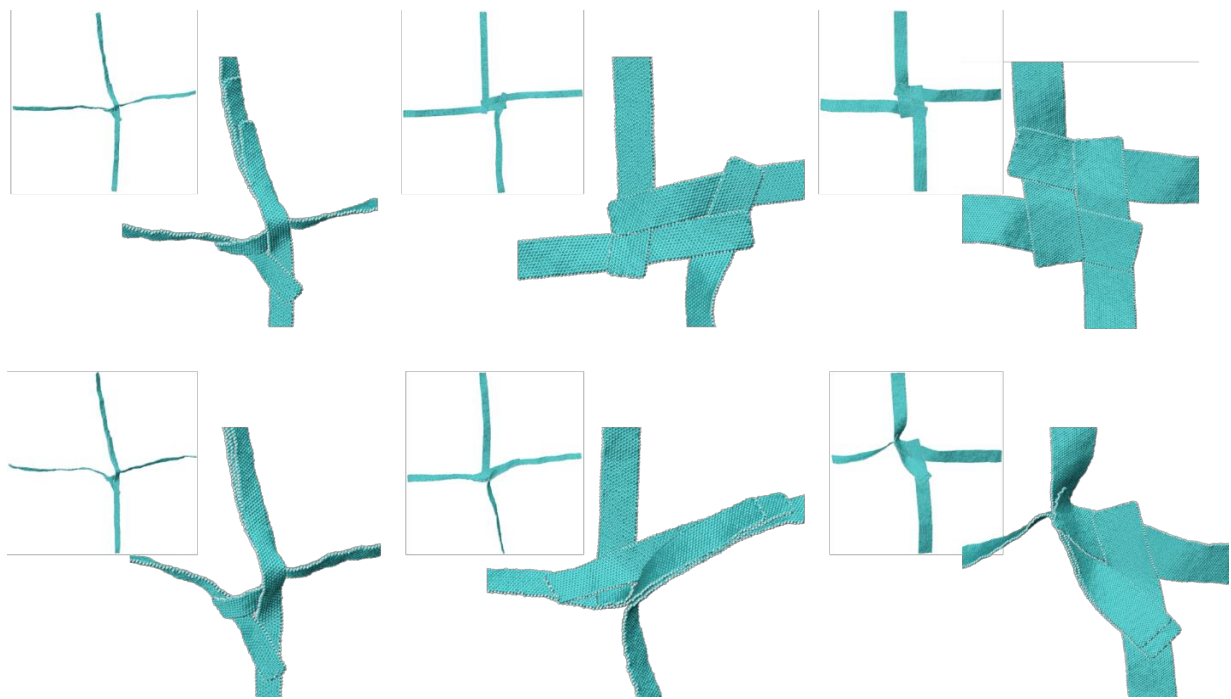

**Figure 5**: Partially equilibrated graphene-based 4-fold RSs made with graphene nanoribbons of 400 Å of length, and 15.1 (left), 25.2 (middle), and 50.4 Å (right) of width, after 1ns (top) and 2 ns (bottom) of MD simulations at 300 K. The insets show the whole structures, and the larger figures show their central parts in order to provide an enlarged view. Cyan (white) colors represent carbon (hydrogen) atoms.

Different from the 3-fold case, the 4-fold RSs did not show much deformation after 100 ps of simulations at 300 K (data not shown). We had to simulate the structures for much longer times. Figure 5, then, shows the partially equilibrated 4-fold RSs after 1ns (top row) and 2 ns (bottom row) of simulations at 300 K. Although we can conclude that 4-fold RSs are also not thermally stable, the 4-fold RSs requires much more time to deform than 3-fold ones. Besides, the amount of deformation of the structures, after a certain time of MD simulation, qualitatively depends on their width as we see Figure 5 from left to right panels. As the 4-fold RSs takes much longer times than 3-fold ones to deform or break the SR conditions, we decided to investigate the effects of two possible constraints: i) reducing the graphene nanoribbon length; and ii) fixing the external extremities of the graphene nanoribbons (i. e., the extremities that are far from the center of the RS). The results are shown in Figure 6 for 4-fold RSs made with graphene nanoribbons of 165 Å of length. For the structures having widths equal 15.1 and 25.2 Å, the graphene nanoribbons bent and twisted towards the increase of the overlap between the surfaces. But for the widest structure tested, even after 3 ns of MD simulation at 300 K, the SR remained. In other words, wide 4-fold RS with their external extremities fixed is thermally stable.

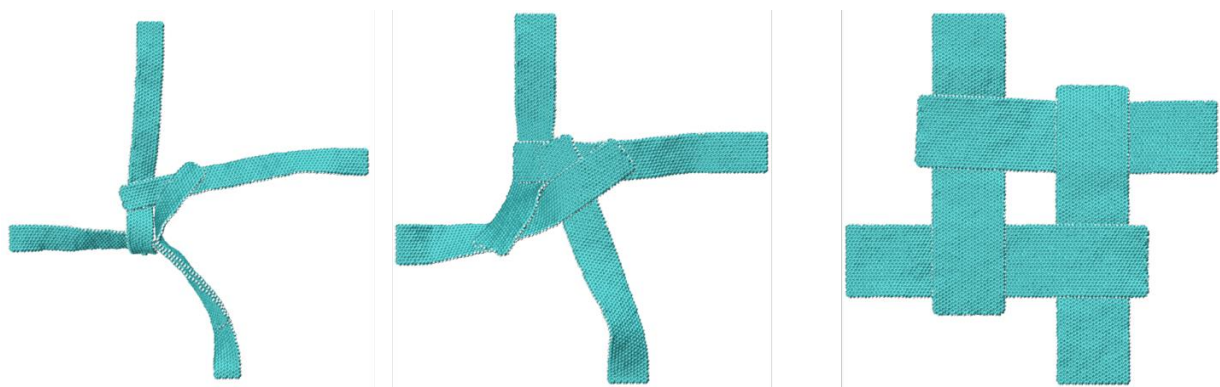

**Figure 6**: Partially equilibrated graphene-based 4-fold RSs made with graphene nanoribbons of 165 Å of length, and 15.1 (left), 25.2 (middle), and 50 Å (right) of width, after 3 ns of MD simulations at 300 K. The external (or far away from the RS center) extremities were kept fixed. Cyan (white) colors represent carbon (hydrogen) atoms.

We also tested another possibility to obtain thermally stable RSs. As some of the RSs studied in the literature have "notches" [1,3], we generated a kind of "notch" by hydrogenating the regions around the contact regions between the graphene nanoribbons. Examples for 3- and 4-fold structures are shown in Figure 3. Results for MD simulations of these RSs are shown in Figure 7. It shows the snapshots of notched 3-fold and 4-fold RSs made with graphene nanoribbons of 400 Å of length and 25.2 Å of width, with fixed external extremities. The magnification of the center of the structures clearly shows that for the 3-fold case, the "notch" did not prevent the structure to change its shape and, again, bending and twisting of the graphene nanoribbons happened compromising its SR. However, even running 5 times longer MD simulations at 300 K of the corresponding notched 4-fold RS, the SR remained. So, we conclude that notched 4-fold RSs are thermally stable. It is important to say that the 4-fold RS with graphene nanoribbons of 400 Å of length and 25.2 Å of width, with fixed external extremities, but without the notch, did not keep the SR after only 1 ns of simulation at 300 K (data not shown).

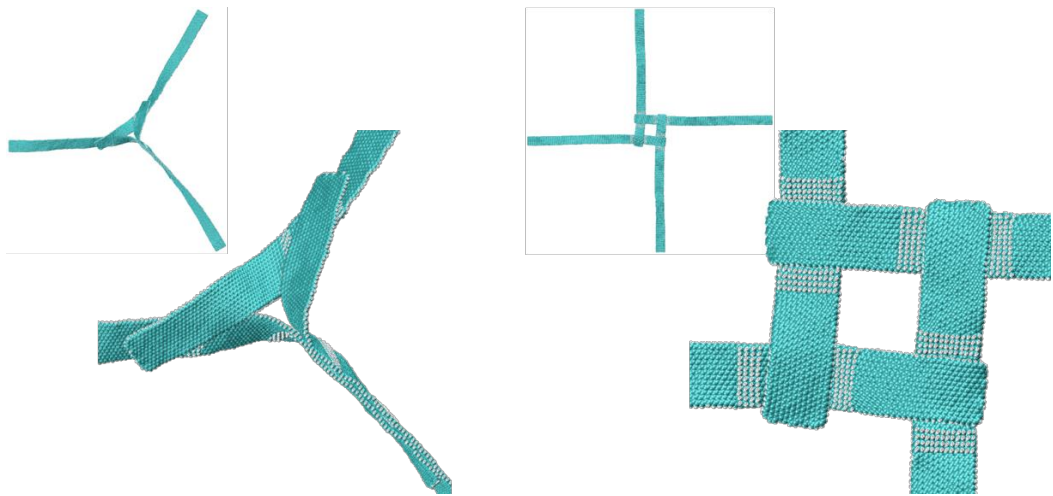

**Figure 7**: Partially equilibrated notched graphene-based 3-fold (left) and 4-fold (right) RSs made with graphene nanoribbons of 400 Å of length and 25.2 Å of width, after 1 ns (left) and 5 ns (right) of MD simulations at 300 K. The external extremities were kept fixed. The insets show

the whole structures, and the larger figures show their central parts in order to provide an enlarged view. Cyan (white) colors represent carbon (hydrogen) atoms.

*3.2. Mechanical impact tests on the thermally stable RSs*

For the thermally stable RSs, the ones shown in the very right side of Figures 6 and 7, the mechanical impact tests were performed. Also, the mechanical test was performed on the corresponding non-RS for comparison and evaluation of the SR effects on the resistance against impact. A mechanical test was also performed on a pristine graphene sample of the same size in order to make an additional comparison of the efficiency of the resistance against impact.

In section 2.2, we described the way the mechanical test was performed. Figure 8 shows the snapshots of the initial and final RSs and graphene structures that were subjected to this test. The corresponding non-RSs were not shown but were also subjected to the same test protocols. In Figure 8, the atoms of the diamond projectile in the final configurations were drawn in transparent color so as to show the pieces of the broken RSs and graphene structures behind it. The energy and the stress of each structure subjected to the mechanical test were collected as functions of their relative positions, i.e., the positions of the center of mass of the structures only. Figure 9 shows the energy as a function of the relative position. One way to quantify the ability of a material to absorb energy by plastic deformation is to calculate its toughness. In order to do that for our structures, we estimated the relative position of the structure where the stress starts growing and took this value as the initial equilibrium length, $L_0$. Then, we computed the strain $\varepsilon = (L - L_0)/L_0$. Figure 10 shows the stress-strain curves of the structures that were subjected to the mechanical test. The stresses were computed by summing up the stress per atom as calculated in LAMMPS [5], so the units are given in units of pressure times volume, or [bar × Å$^3$]. In order to obtain the stress in units of bars or MPa, it would be necessary to compute the atom volume. As the atom volume is ill-defined at this scale, and we are interested only in the comparison of the mechanical performance between RSs and non-RSs, we left the stress in the above units.

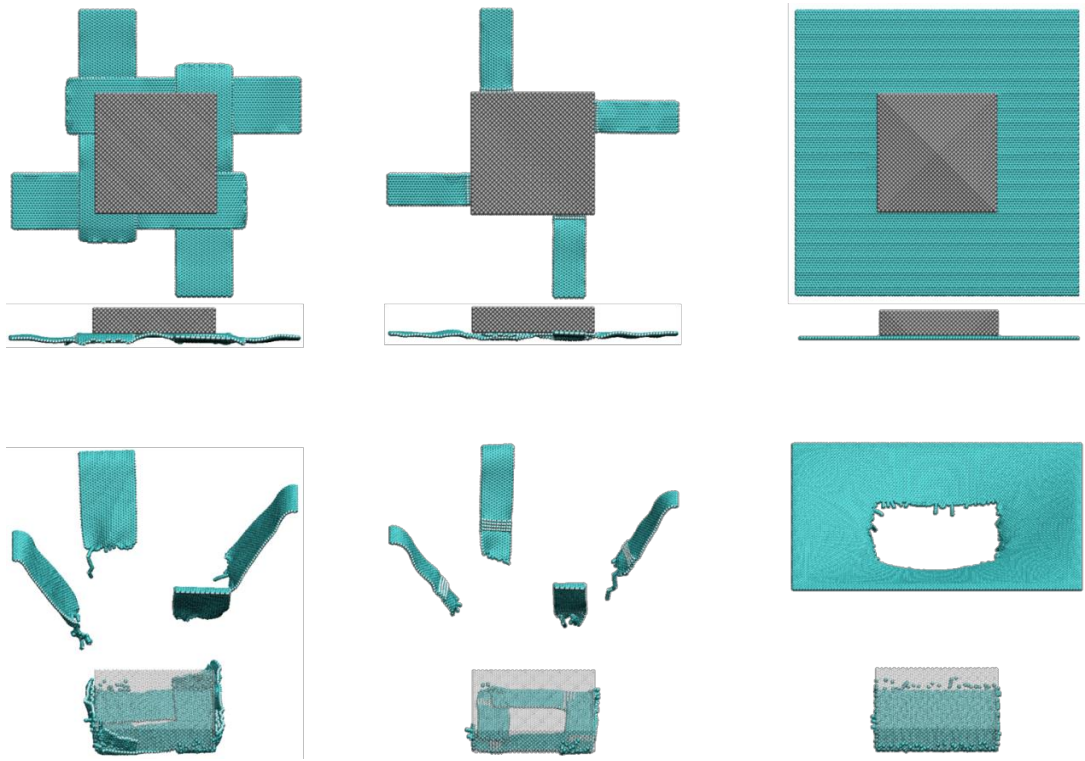

**Figure 8**: Different views of the initial (top: upper and lateral views) and final (bottom: perspective views) structures formed by the diamond projectile of dimensions 100 × 100 × 30 Å and a 4-fold non-notched RS with graphene nanoribbons of 165 Å of length and 50.4 Å of width (left); a 4-fold notched RS with graphene nanoribbons of 165 Å of length and 25.2 Å (middle), and a graphene structure of 240 × 242 Å of size (right). Carbon (hydrogen) atoms of the structures, except the diamond, are shown in cyan (white). Carbon atoms of the diamond are shown in silver (top) and transparent silver (bottom).

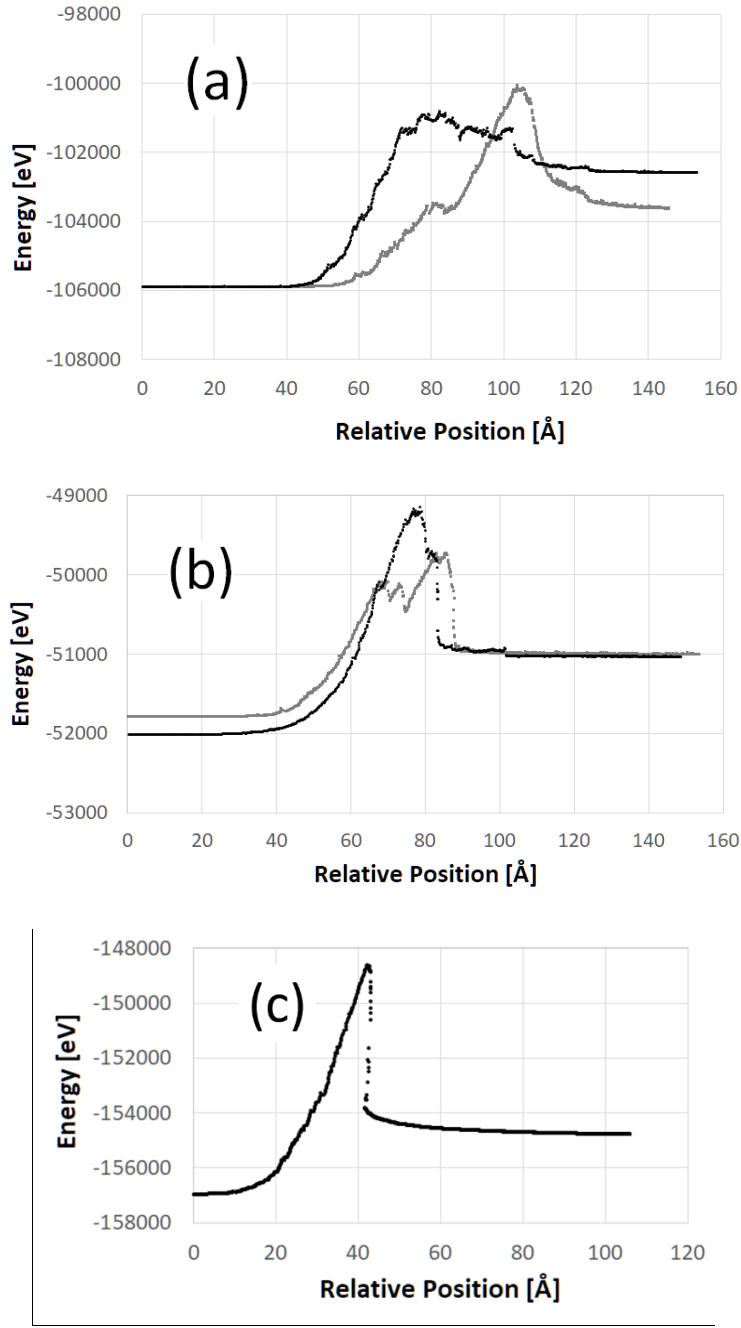

**Figure 9**: Results from the mechanical test. Energy versus relative position of: (a) the 4-fold non-notched RS (black) and non-RS (gray) with graphene nanoribbons of 165 Å of length and 50.4 Å of width; (b) the 4-fold notched RS (black) and non-RS (gray) with graphene nanoribbons of 165 Å of length and 25.2 Å of width; (c) the graphene structure of 240 × 242 Å of size.

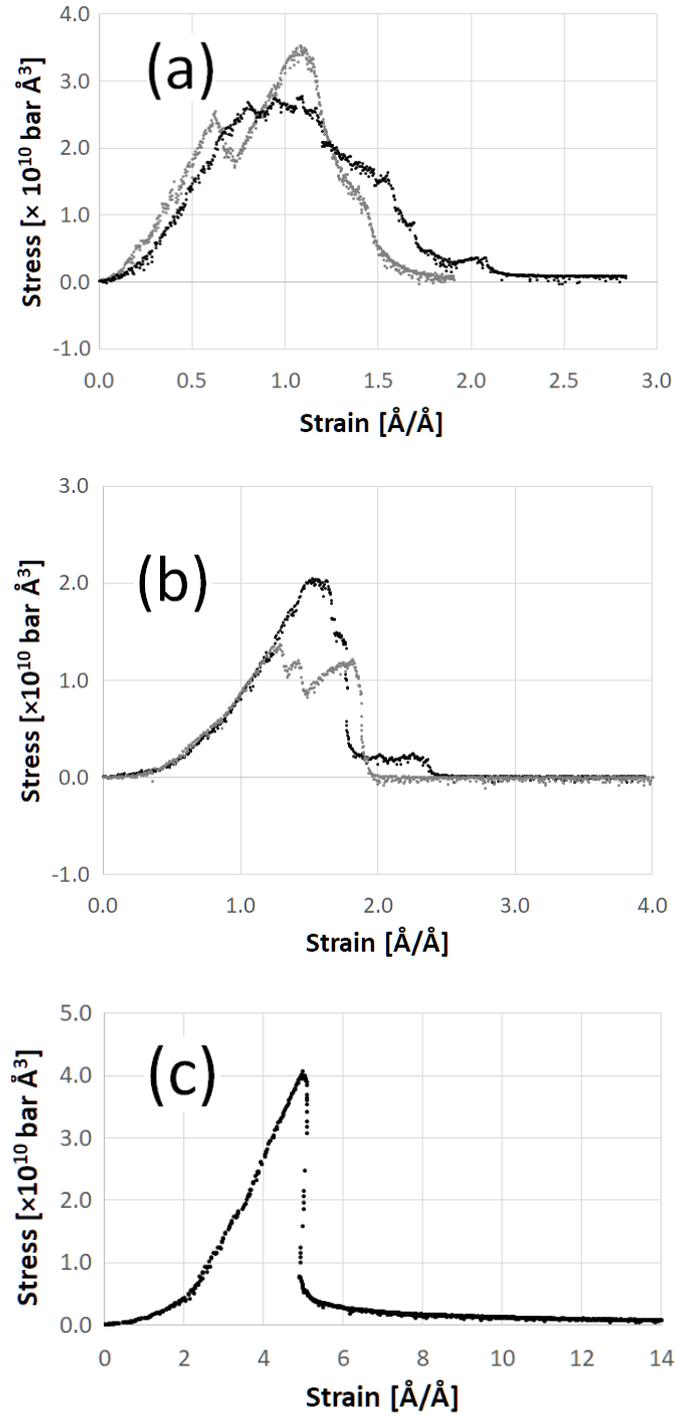

**Figure 10**: Results from the mechanical test. Stress–strain curves for: (a) the 4-fold non-notched RS (black) and non-RS (gray) with graphene nanoribbons of 165 Å of length and 50.4 Å of width; (b) the 4-fold notched RS (black) and non-RS (gray) with graphene nanoribbons of 165 Å of length and 25.2 Å of width; (c) the graphene structure of 240 × 242 Å of size.

Table I below presents the toughness and the differences in the energy between final and initial configurations of all the structures that were subjected to the mechanical test. The toughness was calculated as the area of the stress–strain curves shown in Figure 10.

**Table I**: Toughness, $U$, and herein called absorption energy, $\Delta E = E_f - E_i$, of the structures subjected to the mechanical test. $E_i$ and $E_f$ are the initial and final energies of the structure at the beginning and at the end of the mechanical test, respectively. Units of toughness and energy are [bar Å$^3$] and [eV], respectively.

|  | Non-notched 4-fold RS | Non-notched 4-fold non-RS | Notched 4-fold RS | Notched 4-fold non-RS | Graphene |
|---|---|---|---|---|---|
| $U$ [× 10$^{10}$ bar Å$^3$] | 2.908 | 2.687 | 1.596 | 1.243 | 8.086 |
| $\Delta E$ [eV] | 3300 | 2300 | 1000 | 790 | 2230 |

All these data will be discussed in the next section.

## 4. Discussion

Figure 4 and the right panel of Figure 7 show that none of our 3-fold RSs are thermally stable. The graphene nanoribbons bent and twisted to increase the surface overlapping due to van der Waals interactions, independently on the size of the nanoribbon width. These distortions on the structures compromised the conditions for the SR that is the main object of the present study.

For the 4-fold cases, we also verified the occurrence of bending and twisting of the graphene nanoribbons due to only thermal fluctuations, which also compromised the SR of the structures (see Figure 5). In Figure 5, we showed the snapshots of the 4-fold RSs at two instants of time of different graphene nanoribbon widths simulated without no constraint. However, we also tested the same simulations with the same structures but now imposing a constraint by fixing the external extremities (the extremities that are far away from the center of the RS) of the graphene nanoribbons in order to see if they remain thermally stable. The results (data and structure not shown) showed that the same kind of deformations happened (bending and twisting of the nanoribbons at the center of the RS), with the only difference of taking more time for the structures to get deformed. An interesting observation to notice, however, is that with or without constraints, these 4-fold structures needed much more time (at least 10 times longer than that for the 3-fold cases) of MD simulations to present signs of large deformations. This suggests that there might be some specific configurations or conditions of a 4-fold RS that could lead to thermal stability. In this sense, we found out two different configurations: i) making the 4-fold RS with a shorter graphene nanoribbon (now of 165 Å length); and ii) making a kind of "notch" around the region of contact between graphene nanoribbons (in this case, of 400 Å length). Also, both i) and ii) configurations are subjected to fixing the external extremities of the RS graphene nanoribbons. Figure 6 shows snapshots of 4-fold RSs made with smaller lengthy graphene nanoribbons, and three different values of width (15.1, 25.2 and 50.4 Å, from left to right in Figure 6), simulated at 300 K for about 3 ns. Only the 4-fold RS with 50.4 Å of graphene nanoribbons width maintained the SR conditions in spite of a long time of simulated thermal fluctuations. We considered that this last structure is thermally stable. The second configuration, that of creating a "notch" as shown in Figure 3 by hydrogenating the regions around the graphene nanoribbons contacts, was tested for both 3-fold and 4-fold structures having graphene nanoribbons of 400 Å of length and 25.2 Å of width. The snapshots of 3-fold (4-fold) RSs after 1ns (5ns) of MD simulations at 300 K are shown in Figure 7. It is clear that the 3-fold RS did not keep the SR, but for the 4-fold RS, the notch worked out to prevent the structure to get the deformations that compromise the SR properties of the nanoribbon junctions. This structure is also considered thermally stable.

Now that we identified two thermally stable 4-fold RSs, one notched and another one non-notched, we performed the mechanical impact test with them (see two left panels of Figure 8). Instead of a dynamical impact test as usually performed to study the energy absorption of layered systems [6,7], we designed a quasi-static impact test as described in section 2.2. A dynamical impact test would depend on projectile shape and size, as well as its velocity, and the system we are studying is not a large area, homogeneous layer, but contrarily it is a structure to which we want to have a proof of concept regarding the improvement or not of its mechanical strength due to SR using nanostructures that have low or even none flexural rigidity.

Figures 9 and 11 showed the stress-strain curves and energy versus relative position of the RSs (black curves) and corresponding non-RSs (gray curves) for the mechanical impact tests. The initial and final configurations of the mechanical test with RSs are depicted in the top and bottom rows, respectively, in Figure 8. Also, we performed the test with a pristine graphene sample with the same external dimensions of the other RSs (see right panel of Figure 8) for comparison. The results for the toughness and the differences in the energies of the structures before and after the mechanical impact are displayed in Table I.

Despite the different profiles in the energy versus relative position of the center of mass of RSs (black) and non-RSs (gray) shown in Figure 9, one important result is the comparison of the difference between the final and initial energies between RSs and corresponding non-RSs. These energy differences, $\Delta E$, that will be called here, absorption energies, represent the energy retained in the deformed structures after the projectile passed through them. Although the mechanical test is not dynamical, $\Delta E$ will be taken as measures of the amount of energy the structure absorbs in its own deformations from or due to the impact. Table I shows that the differences in $\Delta E$ between RSs and non-RSs correspond to 26.6% (43.5%) higher for the notched (non-notched) RSs.

The stress-strain profiles of RSs (black curves) and non-RSs (gray curves) shown in Figure 10 are also different. One common point, however, is that all curves started from zero and grew non-linearly suggesting the existence of no linear elastic regime. It comes from the fact that the flexural rigidity of graphene is much smaller than its in-plane rigidity [8]. Therefore, the structures easily bend at the beginning of the movement of the projectile towards them. Only when the carbon-carbon bonds of the structures got tensioned, that the stress within the structures increases. It is important to observe that the graphene nanoribbons that form the RSs are not chemically bonded. So, there is no way to initially tension them. That is why we observe a much larger relative position variation before the energy or stress increase, as compared to the mechanical test of graphene. From the stress-strain curves, we can obtain the toughness of each structure. Table I shows the results of the toughness of all structures and, again, we observe that RSs present larger toughness than their corresponding non-RSs. Both kinds of RSs studied here presented approximated the same value of the difference in toughness, of about $\Delta U \cong 2.21 \times 10^9$ bar Å$^3$ ($\Delta U \cong 3.53 \times 10^9$ bar Å$^3$) for the non-notched (notched) structures, and in favor of the RSs. This is a demonstration of the effectiveness of the structural reciprocity at the nanoscale. Graphene toughness for this test was more than double the highest value that was obtained for the non-notched 4-fold RS with 50.4 Å of width. On the other hand, the absorbed energy, $\Delta E$, for the graphene sample was even smaller than that of non-notched non-RS structure with 50.4 Å of width.

One important issue to be addressed is how a nanostructured RS can be built. To grow precise sizes of graphene nanoribbons and manipulate them has been a challenge. However, recent progress on controlled graphene nanoribbons synthesis [9] combined with the progress towards more precise manipulation [10], suggests that it might be possible to create nanostructured RSs in a near future. For example, the 4-fold structures made of bilayered chromium - aluminum oxide nanoribbons fabricated by Dai *et al*. [11], have structural similarities to our 4-fold RSs.

## 5. Conclusions

In summary, we presented the first computational study of the thermal stability and mechanical response to impact at the nanoscale of graphene-based da Vinci's RSs. 3- and 4-fold RSs. These structures were built based on graphene nanoribbons of different lengths and widths, and tools of molecular dynamics simulations were used to study the thermal stability, at room temperature, equilibrium configuration, and for the thermally stable structures, the mechanical toughness under a projectile impact. Corresponding non-RSs were also studied for comparison. The results showed that 4-fold nanoscale RSs are more resistant than the corresponding non-RSs, even being formed by very pliable bars. These results might be of importance for applications such as nano textiles made of graphene nanoribbons [12]. With the progress on the synthesis of graphene nanoribbons [9] and improvements on their manipulation [10], we believe that it will be possible to experimentally realize the proposed 4-fold RSs.

Regarding the thermal stability, we have observed that for the 3-fold RSs, all structures lost the conditions for structural reciprocity (SR) due to bending and twisting of their graphene nanoribbons under the van der Waals forces that maximized the overlapping between them. The idea of creating a notch to help to keep the graphene nanoribbons in place to maintain the SR was not enough to prevent the whole structure to get deformed. Therefore, we conclude that the 3-fold case was not a good example to verify the advantages of the SR at the nanoscale.

However, for the 4-fold structures, when the length is not so high as compared to the width of the graphene nanoribbons that form them, and if the external extremities are fixed, it is possible to obtain a thermally stable configuration. Also, the idea of creating some kind of notch and keeping their external extremities fixed were shown to be effective to keep the graphene nanoribbons in the right place to maintain the SR, at least, under thermal fluctuations at room temperature.

With the two 4-fold RSs, notched and non-notched, we performed the mechanical test. The results showed that the toughness is larger for RSs than that for non-RSs. This is a demonstration that da Vinci's SR concepts are valid at the nanoscale, even in the limit of the absence of flexural rigidity of the bars that compose the structure. Regarding the effects of defects in the graphene nanoribbons, as long as they did not significantly affect the flexural rigidity of the nanoribbons, we expect the main conclusions to be valid.

One of the applications of RS is related to building resistant bridges, domes, and roofs. If low bending rigidity structures like graphene provides additional mechanical resistance when using SR to make functional structures, stiffer structures like bilayer-bonded graphene [13,14] certainly will allow for building structures like mechanically resistant nano-bridges and nanodomes. This might be of interest for creating isolated regions for reactions or nanoreactors.

We hope the present results will stimulate further studies on the structural reciprocity-properties relationship at the nanoscale.

**Declaration of competing interest**

The authors declare that they have no known competing financial interests or personal relationships that could have appeared to influence the work reported in this paper.

**Acknowledgments**

AFF and DSG are fellows of the Brazilian Agency CNPq (grants 311587/2018-6 and 310052/2019-0, respectively). AFF acknowledges the grant #2018/02992-4 from São Paulo Research Foundation (FAPESP) and from FAEPEX/UNICAMP. DSG acknowledges the Center for Computational Engineering and Sciences at Unicamp for financial support through the FAPESP/CEPID Grant #2013/08293-7. This research also used the computing resources and assistance of the John David Rogers Computing Center (CCJDR) in the Institute of Physics "Gleb Wataghin", University of Campinas.

**Data Availability**

The raw/processed data required to reproduce these findings are available from the corresponding author upon reasonable request.